# NANOSTRUCTURES, MAGNETIC SEMICONDUCTORS AND SPINELECTRONICS

Paata Kervalishvili

Georgian Technical University - 77, Kostava Str., GE 0175, Tbilisi, Georgia

## ABSTRACT

A short overview is given of recent advances in the field of nanosemiconductors, which are suitable as materials for spin polarized transport of charge carriers. On the basis of last theoretical and experimental achievements it is shown that development of diluted and wide forbidden zone semiconductors with controlled disorders as well as their molecular structures is the very prospective way for magnetic semiconductors preparation.

## 1. INTRODUCTION: NANOSCIENCE, NANOTECHNOLOGY AND SPINELECTRONICS.

The main applied instrument of nano science, the interdisciplinary science that draws on physics, chemistry, biology and computational mathematics is a control and manipulation on a nanometre scale, which allows the fabrication of nanostructures with the properties mainly determined by quantum mechanics. Nanostructures constructed from inorganic solids like semiconductors have new electronic and optical properties considerably different from that of the common crystalline state due to their size and quantization effects [1,2]. Quantum dots, for example, have lead to important novel technology for lasers, optical sensors and other electronic devices. The application of nanolayers to data storage, switching, lighting and other devices, can lead to substantially new hardware, for example energy cells, and eventually to the quantum based internet. Nanoscience and nanotechnology encompasses the development of nano spinelectronics, spinelectronics materials production, nano spinelectronic measuring devices and technologies. Nano spinelectronics, based on usage of magnetic semiconductors, represents new area of science and engineering. The reason to that is the perspective of development and creation of principally new materials and devices for information technologies operating as charge, and spin degree of freedom of carriers, free from limitations inherent for metal spinelectronic devices.

## 2. SPIN – POLARIZED TRANSPORT IN SEMICONDUCTORS.

The essential effort of the scientists is concentrated on studying of the spin-polarized transport in nanosize multilayer structures, which are including alternating layers of ferromagnetic metals and non-magnetic semiconductors. Operation of a spintronic device requires efficient spin injection into a semiconductor, spin manipulation, control and transport, and spin detection. The relevant role in solution of this problem is shunted to search and investigations of new ferromagnetic materials, which are capable to be reliable and good spin injectors. Among such objects the magnetic discrete alloys are very promising. They are multilayer systems composed of submonolayers of a ferromagnetic material in the matrix of a non-magnetic semiconductor, for example, Mn/GaAs or Mn/GaSb. It is well known, that these alloys have high Curie temperatures and sufficiently high spin polarization. The circumstance is not less important that it is possible to control and to manage of the "ferromagnetic metal - semiconductor" boundary surface immediately during the synthesis of these materials.

As it was investigated recently they should be prepared only by the methods of the MOS hydride epitaxy and laser epitaxy with usage of pulsed annealing of epitaxial layers. Perspectives of development and creation of new types of a non-volatile memory stipulate the significance of spintronics with random access (MRAM), quantum single-electron logical structures, and ultra dense information storage media. Thus, elementary information storage unit will be represented by an electron spin. The realization of the spin-polarized current transfer opens out new possibilities for the solid-state electronics also. For instance, there are observations of the spin-polarized luminescence and creation of the high frequency diodes, output characteristics which ones one can change by an external magnetic field. Another example is the possibility for creation of a new generation of narrow-band devices of the solid-state electronics of millimetre and submillimeter wave ranges like generators, amplifiers, receivers and filters, modulated and frequency tuned by magnetic field and fully current controlled [3].

The discovery of giant magnetoresistance effect (GMR) can be considered as new important step for development of spintronics. This phenomenon is observed during the study of thin films with alternating layers of ferromagnetic and non-magnetic metals. It is found that, depending on the width of a non-magnetic spacer, there can be a ferromagnetic or antiferromagnetic interaction between magnetic layers, and antiferromagnetic state of magnetic layer can be transformed in ferromagnetic state by an external magnetic field. The spin-dependent scattering of conduction electrons is minimal, causing a small resistance of material, when magnetic moments are aligned in parallel, whereas for antiparallel orientation of magnetic moments the situation is inversed. The effect GMR brightly has demonstrated, that a spin-polarized electrons can carry magnetic moment through non-magnetic materials.

Namely the GMR effect was used in a new generation of the magnetic field sensors, which appeared in 1994 as commercial products on market. But present boom in industry producing the information storage devices started a bit later, in 1997, when the IBM Company has presented the first hard drives with the GMR reading heads. The implantation of this technology has allowed more than on the order to increase a density of the information storage on magnetic disks. The sensors operating with the tunnel magnetic junctions (MTJ) fall into the second class spintronics devices. Very thin dielectric layer divides their ferromagnetic electrodes, and electrons are tunnelling through a nonconducting barrier under influence of applied voltage. The tunnel conductivity depends on relative orientation of the electrode magnetizations, and tunnel magnetoresistance (TMR) it is small for parallel alignment of magnetizations of electrodes and is high in opposite case. New memory devices include storage units based on the MTJ structures and allow not only to increase essentially the storage density and the access speed to a memory, but also to provide complete saving of data at disconnecting of a power supply. A disadvantage of these devices is the small scale of integration, bound with necessity of usage of additional controlling transistors. Possibility to overcome these limitations is connected nowadays only with the development of the semiconducting spintronics, and, in particular, with the creation of spin transistors. In this case spintronic devices cannot only switch or to detect electrical and optical signals, but also to enhance them, and also to be used as multifunction units. Due to this reason, the third direction of development spintronic devices is based on the development of multilayer nano structures of ferromagnetic semiconductors, which demonstrate properties not available for their metal analogs. One can refer to number of these properties the possibility to control by electric field a magnetic state of material and the giant planar Hall effect, which exceeds on several orders of magnitude the Hall effect in metal ferromagnets. The super-giant TMR effect observed for the first time in epitaxial (Ga,Mn)As/GaAs/(Ga,Mn)As structures is not less promising for applications.

There are no effective ways of injection the spin-polarized current in non-magnetic semiconductors at the present moment. The spin injection from magnetic semiconductors in non-magnetic gives good results in a number of cases, but while it has a place only at low temperatures, far from room temperature.

So-called magnetic discrete alloys to days are of the most prospective materials for solution of the spin injection problem. These alloys involve a periodic system of sub-monolayers of magnetic ions (for example, Mn), placed between semiconducting layers (GaAs, GaSb, InAs) forming a magnetic superlattice. There are as incidentally distributed Mn ions and 2D magnetic islands of MnAs (or MnSb) as well in manganese containing layers. The discrete alloys have high Curie temperatures (above 300 K for the GaSb-system), demonstrate extraordinary Hall effect at high temperatures and have a relatively high degree of the spin polarization. It is possible in such systems to control not only quality of the border "ferromagnetic metal - non-magnetic semiconductor", but also manage of the current carrier's concentration and change the type of magnetic ordering. The discrete alloys should be considered as random magnet systems owing to hardly inhomogeneous allocation of a magnetic phase in sub-monolayers.

The interest in so called diluted magnetic semiconductors is given an impetus by the recent demonstration of the ferromagnetic critical temperature $T_c$ =110 K in GaMnAs. To date, most theoretical models proposed assume that the holes occupy a Fermi sea in the valence band [4].

Theoretical models of the virtual crystal approximation have been used to study the influence of disorder on transport and magnetic properties of magnetic semiconductors. The Boltzmann equation with Born approximation scattering rates has provides estimates of Anizotropic Magneto resistance Effect of order up to 12 %. The key of understanding the kinetic and magnetic anisotropy effects is a strong spin-orbit coupling in the basic semiconductor valence band.

The most striking feature in off-diagonal conductivity coefficients for example in (GaMn)As and other arsenide and antimonide of diluted magnetic semiconductors occurs is the large anomalous Hall effect, which occurs because of spin-orbit interactions. In the metals standard assumption is the Anomalous Hall arises because of spin-orbit coupling component in the interaction between band quasiparticles and crystal defects, which can lead to skew scattering with Hall resistivity contribution proportional to diagonal resistivity. For diluted magnetic semiconductors the Anomalous Hall effect is based on spin-orbit coupling in the Hamiltonian of the ideal crystal and implies a final Hall conductivity even without disorder. The Hall fluid effective Hamiltonian theory discussed without free parameters and it is reliably in materials with high critical temperatures where halls are metallic.

The effects of the $As_{Ga} - As_i - V_{Ga}$ transition to the ferromagnetism of (GaMn)As can be explained by the Mulliken orbital populations of the d-shell for both majority and minority spins and the corresponding spin polarization for the ferromagnetic configuration. In this case the ferromagnetic coupling is strengthening considerably by the distortion, and that all together the energy splitting and Mulliken orbital population of $As_i$ $V_{Ga}$ are the very similar to those of defect free (GaMn)As. These suggest that the ferromagnetic order in (GaMn)As is unaffected by the presence of $As_i$ $V_{Ga}$ pairs. This result is in agreement with the hole-mediated picture of ferromagnetism, and can understood by noting that $As_i$ $V_{Ga}$ defects energy levels show minimal splitting in (GaMn)As [5].

More detailed studies of disorders will combine Kondo description of the spin interactions as well as relevant Monte Carlo techniques applied to both metallic and insulating conditions.

## 3. NANOSTRUCTURES OF WIDE FORBIDDEN ZONE SEMICONDUCTORS.

High-temperature semiconductors with wide forbidden zones are also the very promisable materials for modern nano electronics. Materials based on carbon and boron provides complicated substances with unique structural properties. The technology for their film preparation is promising with their desirable electric and physical properties such as mechanical hardness and chemical resistance. Research conducted during the last decades of the 20th century have shown that carbon and boron crystals form clusters, the essential structural elements of which contain 4, 12, 60, or 84 atoms. These nanoelements, due to their thermodynamic properties, transform to amorphous or crystalline films, layers and other deposits, which have some advanced properties.

The clusters having a stable configuration under equilibration conditions take the forms of different geometrical figures - from triangular to dodecahedral and icosahedrical.

According to the classical idea of particle formation and growth and in correspondence with the so-called atomistic process of conception, atoms being the germ of the solid phase unite in aggregates (clusters) where their quantity is dependent on their atomic potentials.

Statistical calculations of the thermodynamic properties of small clusters carried out by means of computer modelling have shown that the potential energy of the atomic cluster components is the main factor determining the chemical potential of the cluster.

The growth in the quantity N of atoms in the cluster results in the increase of the thermodynamic potential P (N), caused by the increase in atoms at the surface. At the same time, the increase of surface energy accompanying the additional atoms is not continuous, but discrete because of the differences between the energetic contributions of the atoms completing the formation of the co-ordinating sphere [6].

Further growth in the aggregate leads an increase in the volume by means of a gradual addition of atoms from the sides to the growing cluster - volume growth. Using the established and recent approach to the mechanism of cluster formation, it is easy to show that the appearance of small particles analogous to the so-called fractal clusters very often takes place. Following this, the particle growth occurs not by the joining of separate atoms to their existing aggregate, but by a conglomeration of aggregates with stable configuration, which preserves their individual properties. The formation of small particles (clusters) is actually carried out by various methods, among which are supersonic outflow of vapours into the vacuum, thermo-, laser- and plasma-chemical modes of substance reduction from their gas-phase compounds, vapour precipitation upon cold substrates, reaction of molecular effusion from a cell, and etc. These techniques are being used to study the process of the small particle formation, volume growth and growth on specially prepared surfaces.

The production of elementary boron is presently being developed by various powder and film technologies. The greatest interest is with modes of small particle production to provide high dispersion and purity as well as the study of the processes of cluster conception and growth.

Established theory and experiment have shown that the elementary boron atoms group into an aggregate of icosahedrical form consisting of 12 boron atoms ($B_{12}$). Usually the boron small particles consist of one or more icosahedrons united in a cluster or various configurations depending on the thermodynamic conditions at formation.

Designating the chemical potential of the structural element (in boron this is a 12-atom icosahedron, in carbon it is a 4-atoms tetrahedron) as E, and the chemical potential of the flat particles (cluster) as P, it is apparent that an equilibrium between the longitudinal dimensions and flat cluster thickness will be achieved when

$m_i E - m_i P = 2\alpha v/r$, where $\alpha$ is the surface specific free energy per one structural element, v is the specific volume of cluster per one structural element (this is analogous to the Gibbs-Thomson expression), r - radius vector . Given that the equilibrium form is subordinated to the second order non-linear differential equation [7] and the difference $m_i E - m_i P$ is constant over the whole surface of a particle, the solution of this equation represents the envelope of the cluster:

$$\vec{n} \cdot \vec{r} = 2\alpha (\vec{n}) v /( m_i E - m_i P)$$

where $\vec{n}$ is the vector of the normal to the envelope ( the surface) of the small particle as determined by the radius vector $\vec{r} =$ r (the expression is analogous to the Curie-Wulf formula. From this equation it is possible to evaluate the geometric form and longitudinal dimension of the equilibrium state of the cluster of elementary boron in flat form (fig.1) under given thermodynamic conditions.

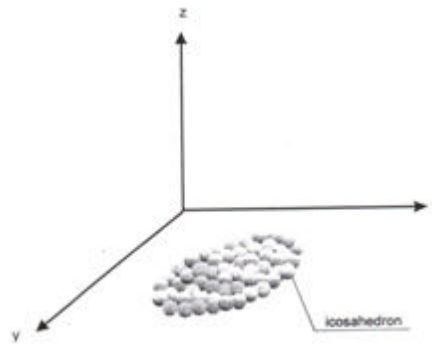

Fig.1 - Schematic picture of a cluster of 12 - atom icosahedrons in plane form

During electron-microscopic studies and testing of the structure of elementary boron produced by means of boron three-chloride reduction with hydrogen and laser-chemical multi-photon dissociation of the dichlorboran molecule, the observed structural elements - boron icosahedrons, are statistically disseminated in an amorphous condition and in a crystalline condition with rhombohedric symmetry.

During the electron-microscopic study with the electrons acceleration voltage of 100 kV, light-permissible pictures were produced with microscopic magnification $2 \times 10^{-5}$. The electron microscopy pictures show that the particle has an amorphous structure, boron icosahedrons are placed non-regularly in the plane [7].

The analysis of the micro-diffractional picture, consisting of three diffusion halos, as well as calculations of the interplane spaces has given values consistent with those for the planes of the icosahedron $B_{12}$. A number of boron particles partly overlapping each other is shown in fig.2. The particles form plane structures with longitudinal dimension 20-40 times as large as their thickness. The thickness is approximately equal to the linear size of 12-atoms of boron.

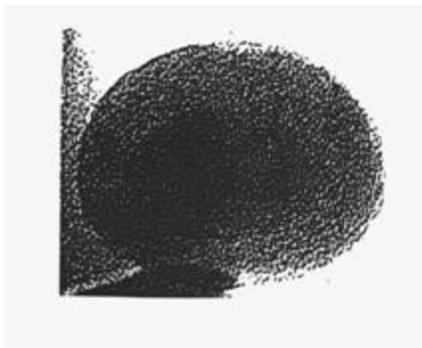

Fig. 2 - Electron microphotograph of boron particles which are overlapping each other

The analysis of the analogous electron microphotographs shows that the increase of the longitudinal dimensions of the particles under the thermal treatment occurs because of the joining of lammelar clusters formed from boron icosahedrons with the initial dimension 2-5 nm and icosahedron thickness, and because of their stratification.

The space between the normal stripes is 0,8 nm, which is very near to the interplane space of planes (111) of β - rhombohedrical boron ($d_{111}$ =0.7962 nm) in crystalline structure and coincides with values calculated from interreflex space data.

Thus, the direct observation of small particles of elementary boron using high-permissible electron microscopy shows that the boron clusters (2-5 nm) are the amorphous plane compounds where the ratio of the thickness to the longitudinal dimension varies from 1:10 to 1:40. These clusters which consist of non-regularly thermally treated small particles of boron proceed to crystallisation, which at first occurs in the centre of a particle without its plane structure, then advances to the stage of a partly crystallized clusters, stratification and finally volume crystallisation.

The elementary particles of boron produced by the plasma-chemical method in the free-poured state have shown an unknown effect – the appearance of a plant-shape cluster. In other words, the ultra-dispersive amorphous boron powder clusters consisting of statistically (nonregular) distributed icosahedrons has been found. The thermal treatment of the elementary boron powder, which consists of plane clusters, in a vacuum furnace as well as in the electron microscope's beam resulted at first in a transition of icosahedrons from a statistical into a modulated condition and then their grouping into a volume rhombohedrical configuration of crystalline boron of β-type. At the first stage there has been observed a regulation of the icosahedrons in the flat form (fig.3), and then at the second stage a unification of these forms accompanied by the crystalline phase formation.

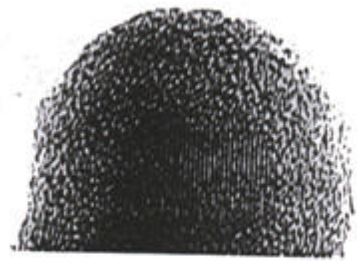

Fig. 3 - Electron microphotograph of a small particle of boron at the first stage of regulation

Similar results were observed in the case of ultra-dispersive carbon powders. The electron microscope studies show that the carbon powder consists of small structural elements in the forms of tetrahedral disks. On receipt of a very small portion of energy they organized into the bigger clusters, sometimes the so-called Fullerenes. In the electron microscope beam after treatment they become crystalline structures and in suitable thermodynamic conditions, become diamond.

The same situation occurs with carbon layers prepared by the laser spraying - laser plasma technologies [9].

The Laser Spraying Method for film and layer preparation and its most attractive and advanced development, Laser Plasma Deposition, was used for boron and carbon thin film and layer production. Investigations have shown that the prepared crystalline layers have a diamond like structure with the lattice parameters close to that of crystalline diamond [10].

Further study of small particles, their conception and growth will explain a number of natural phenomenon of the formation of small solid structures.

The practical importance of this investigation consists of the possibility to create new technologies for the production of ultra-dispersive materials with given crystalline or amorphous structure and the requisite properties. Recent experimental studies of small solids - particles, structural elements of some non-organic (carbon, cobalt, etc.) and biological systems (biomolecules, bio-solids) have brought new data regarding the nature of the kinetics of their formation.